\documentclass[article,twocolumn,
amsmath,amssymb,
aps,
prl,
floatfix,
]{revtex4-2}

\usepackage{float}
\usepackage{graphicx}
\usepackage{dcolumn}
\usepackage{bm}
\usepackage{hyperref}
\hypersetup{
    colorlinks=true,
    linkcolor=blue,
    filecolor=magenta,      
    urlcolor=cyan,
}
\usepackage{color}
\usepackage{fancyhdr}
\usepackage[normalem]{ulem}

\newcommand{\risp}[1]{\textcolor{black}{{#1}}}

\usepackage{tikz,xcolor,hyperref}

\definecolor{lime}{HTML}{A6CE39}
\DeclareRobustCommand{\orcidicon}{
	\begin{tikzpicture}
	\draw[lime, fill=lime] (0,0) 
	circle [radius=0.16] 
	node[white] {{\fontfamily{qag}\selectfont \tiny ID}};
	\draw[white, fill=white] (-0.0625,0.095) 
	circle [radius=0.007];
	\end{tikzpicture}
	\hspace{-2mm}
}
\foreach \x in {A, ..., Z}{\expandafter\xdef\csname orcid\x\endcsname{\noexpand\href{https://orcid.org/\csname orcidauthor\x\endcsname}
			{\noexpand\orcidicon}}
}

\begin{document}

\title{Implosive Dynamics from Topological Quenches in Bose-Einstein Condensates}

\author{Marios Kokmotos\orcidC{}}
\affiliation{School of Physics and Astronomy, University of Birmingham, Edgbaston, Birmingham, B15 2TT, United Kingdom}
\author{Dimitri M. Gangardt\orcidB{}}
\affiliation{School of Physics and Astronomy, University of Birmingham, Edgbaston, Birmingham, B15 2TT, United Kingdom}
\author{Giovanni Barontini\orcidA{}}
\email{g.barontini@bham.ac.uk}
\affiliation{School of Physics and Astronomy, University of Birmingham, Edgbaston, Birmingham, B15 2TT, United Kingdom}

\date{\today}

\begin{abstract}
We show numerically that a repulsive Bose-Einstein condensate can be driven into implosive dynamics by a direct topological quench. We first realize giant vortices by quasi-adiabatic phase imprinting, and then perform a sudden anti-imprint that cancels the accumulated winding in a single step, abruptly switching the condensate from a highly charged vortex state to the trivial sector. The resulting phase-density mismatch launches a rapid inward radial flow and produces a strong central density buildup, despite the repulsive interactions.  After the first implosion, the dynamics evolves into circular nonlinear wave fronts that subsequently undergo breaking of azimuthal symmetry (axisymmetry) down to a polygonal one, whose shape is determined by the way the giant vortex is built. These results establish topological engineering as a new tool for studying implosive dynamics and symmetry-breaking instabilities in quantum fluids.
\end{abstract}

\maketitle

Collapse is a recurrent theme across nonlinear physics: when an equilibrating
``pressure'' term is suddenly removed or overwhelmed, an initially stable
configuration can undergo rapid self-focusing and reach extreme densities.
In self-gravitating matter, collapse is triggered once thermal or degeneracy
pressure can no longer support the mass; in particular, electron-degeneracy
support in white dwarfs fails above the Chandrasekhar mass
\cite{Chandrasekhar1931}, while loss of pressure support in massive stellar
cores underlies core-collapse supernovae \cite{Bethe1990}. Rotation can delay,
reshape, or transiently halt this process via centrifugal support
\cite{Dimmelmeier2008}.

In many nonlinear media, a central question is how an initially radial focusing, collapsing or converging event subsequently loses axisymmetry and selects low-order azimuthal modes. Polygonal or symmetry-broken wave patterns are a recurrent manifestation of this broader phenomenology, appearing in a range of rotating, annular, and converging flows, e.g. hydraulic jumps, rotating free-surface flows, and Rossby-wave or jet-instability patterns such as Saturn's north-polar hexagon \cite{Bush2006,Jansson2006,SanchezLavega2014}. Implosive systems are likewise often highly sensitive to low-mode asymmetries \cite{Izumi2021}. In these contexts, a central issue is to determine how symmetry breaking is seeded, selected, and amplified during and after the focusing stage.

Dilute Bose-Einstein condensates offer a clean and controllable
setting in which to address this question. Their mean-field dynamics is governed
by a nonlinear Schr\"odinger equation common to many nonlinear wave systems,
while both the interaction strength and the external confinement can
be tuned with high precision \cite{Dalfovo1999}. In repulsive condensates,
strong density gradients and compressive flows are known to generate quantum
shock waves and dispersive shock fronts rather than classical dissipative shocks
\cite{Dutton2001,Hoefer2006,wqPhysRevLett.125.180401}. Their circular counterparts, ring dark solitons,
provide a natural route from radial focusing to symmetry breaking: in higher
dimensions they are prone to snake instability and can decay into
vortex-necklace-like states once rotational symmetry is broken
\cite{Theocharis2003,Toikka2013}. Related angular-mode selection also appears in
the splitting of giant vortices, where the number of fragments is tied to the
angular momentum of the unstable mode \cite{Kuopanportti2010}. Here we show, using numerical simulations, that implosion dynamics and subsequent polygonal symmetry breaking can be engineered topologically in Bose-Einstein condensates. 

\begin{figure*}
\centering
\includegraphics[width=0.9\textwidth]{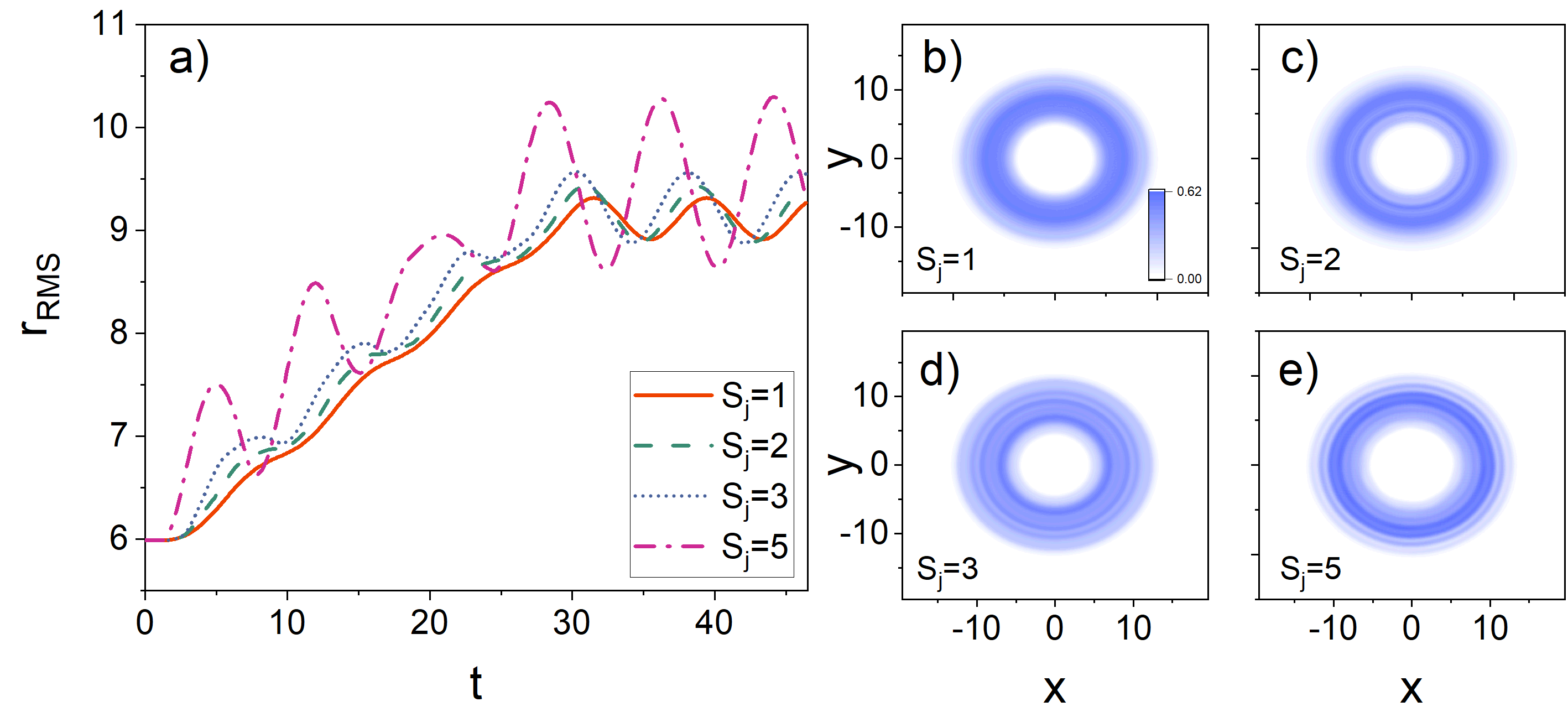}
\caption{Formation of a giant vortex with final winding $S_0=25$ using sequential topological imprints. a) Time evolution of the RMS radius applying different step sizes $S_j=\{1,2,3,5\}$, while keeping the imprinting slope $S_j/dt$ fixed. b)-e) Corresponding column-density profiles at the first maximum of the breathing cycle for $S_j=1$, $2$, $3$, and $5$, respectively.}
\label{Fig1}
\end{figure*}

Quantized vortices
are topological excitations characterized by an integer winding number
\begin{equation}
S_0=\frac{1}{2\pi}\oint_{\mathcal C}\nabla\varphi\cdot d\boldsymbol{\ell}\in\mathbb{Z},
\label{eq:winding}
\end{equation}
where \(\varphi\) is the condensate phase and \(\mathcal C\) encloses the defect
core. A giant vortex, characterized by large \(S_0\), supports an annular
density profile through its azimuthal superflow
$u_\theta(r)={\hbar S_0}/{m r}$, with $m$ the atomic mass and $r$ the radial coordinate,
which provides centrifugal support \cite{Fetter2009}. multi-quantized vortices are typically energetically and
dynamically unstable and tend to split into single quantized vortices
\cite{Shin2004,Kuopanportti2010}. Long-lived giant-vortex cores can nevertheless
be realized under rapid rotation and/or with anharmonic or pinning potentials
\cite{Engels2003,Kasamatsu2002,Lundh2002,Simula2004}.

Here we engineer giant vortices by sequential quasi-adiabatic phase imprinting
\cite{easton2023vortex,Leanhardt2002} in a highly oblate Bose-Einstein condensate, and then
perform a \emph{topological quench} that cancels the accumulated winding, implementing
an abrupt switch between distant topological sectors. This direct topological
quench is enabled by the quantum nature of the superfluid: because circulation
is quantized, phase anti-imprinting can deterministically switch the system
between topological sectors, an operation with no counterpart in an ideal
classical fluid. The subsequent dynamics is caused by the removal of the
azimuthal flow on a timescale that is short compared with the hydrodynamic
response. The annular density profile cannot relax instantaneously and,
deprived of centrifugal support, it undergoes a rapid implosive focusing even
though the underlying interactions remain repulsive. Beyond the initial
refocusing, the dynamics develops circular nonlinear wave fronts that
subsequently break axisymmetry into polygonal patterns. Crucially, the outcome
is controlled not only by the final winding \(S_0\), which sets the strength of
the implosion, but also by the preparation protocol itself: the imprinting step
\(S_j\) seeds the azimuthal symmetry content that is later amplified during the
post-quench breakup.

Our results therefore establish a route to \emph{topological engineering of
symmetry-breaking implosions} in a repulsive quantum fluid. They identify
topological preparation and topological quench as complementary non-equilibrium
controls: the former imprints the symmetry content of the state, while the
latter releases the stored circulation and launches the implosion. This opens a
controlled route to investigate how selected
low-order asymmetries shape implosive dynamics in systems where such
asymmetries are otherwise difficult to seed and control
\cite{Bush2006,Jansson2006,Izumi2021}.

We numerically integrate the time-dependent Gross-Pitaevskii equation (GPE)
in dimensionless form,
\begin{equation}
i\,\partial_t \Psi(\mathbf{r},t)=\left[-\frac{1}{2}\nabla^2+V(\mathbf{r})
+g\,|\Psi(\mathbf{r},t)|^2\right]\Psi(\mathbf{r},t),
\label{eq:gpe_dimless_methods}
\end{equation}
where the trapping potential $V$ is
harmonic with aspect ratio \(\lambda=\omega_z/\omega_\perp\gg 1\). Length and time are measured in harmonic oscillator units \(a_{ho}=\sqrt{\hbar/(m\omega_{ho})}\) and \(\omega_{ho}^{-1}\), with \(\omega_{ho}=\lambda^{1/3}\omega_\perp\), and \(m\) the
atomic mass. The interaction coefficient is \(g=4\pi a_s\), with \(a_s\) the \(s\)-wave scattering length. 

For concreteness, we consider \(N=5\times10^4\) \(^{87}\)Rb atoms with in an oblate trap with \(\lambda=15\)
and \(\omega_\perp=2\pi\times 20\,\mathrm{Hz}\). For these parameters the scattering length is \(a_s\simeq0.0034\), while the healing length is \(\xi\simeq 0.2\), the condensate is in the Thomas-Fermi regime $Na_s\gg1$, and the dynamics is effectively two-dimensional. We solve Eq.~\eqref{eq:gpe_dimless_methods}
on a \(512\times512\times16\) grid with a resolution of $0.075\times0.075\times0.12$. Time propagation is performed with a split-step Fourier method with timesteps of 1.55$\times10^{-4}$. 

Vorticity is generated by phase imprinting. At an imprint time \(t_j\) we apply an
instantaneous phase mask \cite{easton2023vortex},
\begin{equation}
\Psi(\mathbf{r},t_j^+)=\Psi(\mathbf{r},t_j^-)\,\exp\!\left[i\,S_j\,\theta(x,y)\right],
\label{eq:imprint_methods_simple}
\end{equation}
where \(S_j\in\mathbb{Z}\) is the imprinted winding number and
\(\theta(x,y)=\arg(x+iy)\) is the polar angle about the trap centre. This corresponds
to imprinting a multi-quantized vortex of charge \(S_j\) in the center of the condensate. \risp{Such phase profiles can be generated experimentally using spatially shaped optical potentials produced by spatial light modulators or digital micromirror devices, which imprint the required phase onto the condensate through the AC Stark shift \cite{Dobrek1999,meyer}.}

The strong confinement along \(z\)
suppresses vortex bending, so the dynamics is effectively two-dimensional over the
timescales of interest. Directly imprinting a large-\(S_0\) giant vortex in a single step is highly disruptive:
the abrupt injection of a strong azimuthal phase gradient 
drives violent density depletion and rapid fragmentation of the vortex core. Instead, here we
construct giant vortices stepwise by applying a sequence of imprints that increase the
charge as adiabatically as possible (given the quantized nature of the process). Between imprints the condensate is evolved under
Eq.~\eqref{eq:gpe_dimless_methods}, allowing the density to adjust and the central hole
to expand. This enables us to reach large total winding numbers $S_0$ while preserving a coherent superfluid. We have tested this method with sequential imprints of $S_j=\{1-7\}$, and with different times between imprints, finding it robust across a wide parameter range. In our simulations we have reached vortices with $S_0=25$, limited only by the size of the numerical grid.

In Fig.~\ref{Fig1} we show representative realizations of a final giant vortex with $S_0=25$, obtained with different imprinting steps $S_j$, while keeping the imprinting rate $S_j/dt$ fixed. Panel~a) reports the corresponding evolution of the RMS radius. In all cases the protocol excites a radial breathing mode; however, as expected, increasing $S_j$ produces oscillations of larger amplitude. Panels~b)-e) show the corresponding column density profiles at the first maximum of the breathing cycle. Besides the overall breathing motion, larger imprinting steps also generate more pronounced concentric density ripples, consistent with additional phonon-like radial excitations. Nevertheless, the method remains robust and reliably produces a coherent giant vortex even for the largest imprinting steps considered here. 

\begin{figure}
\centering
\includegraphics[width=0.45\textwidth]{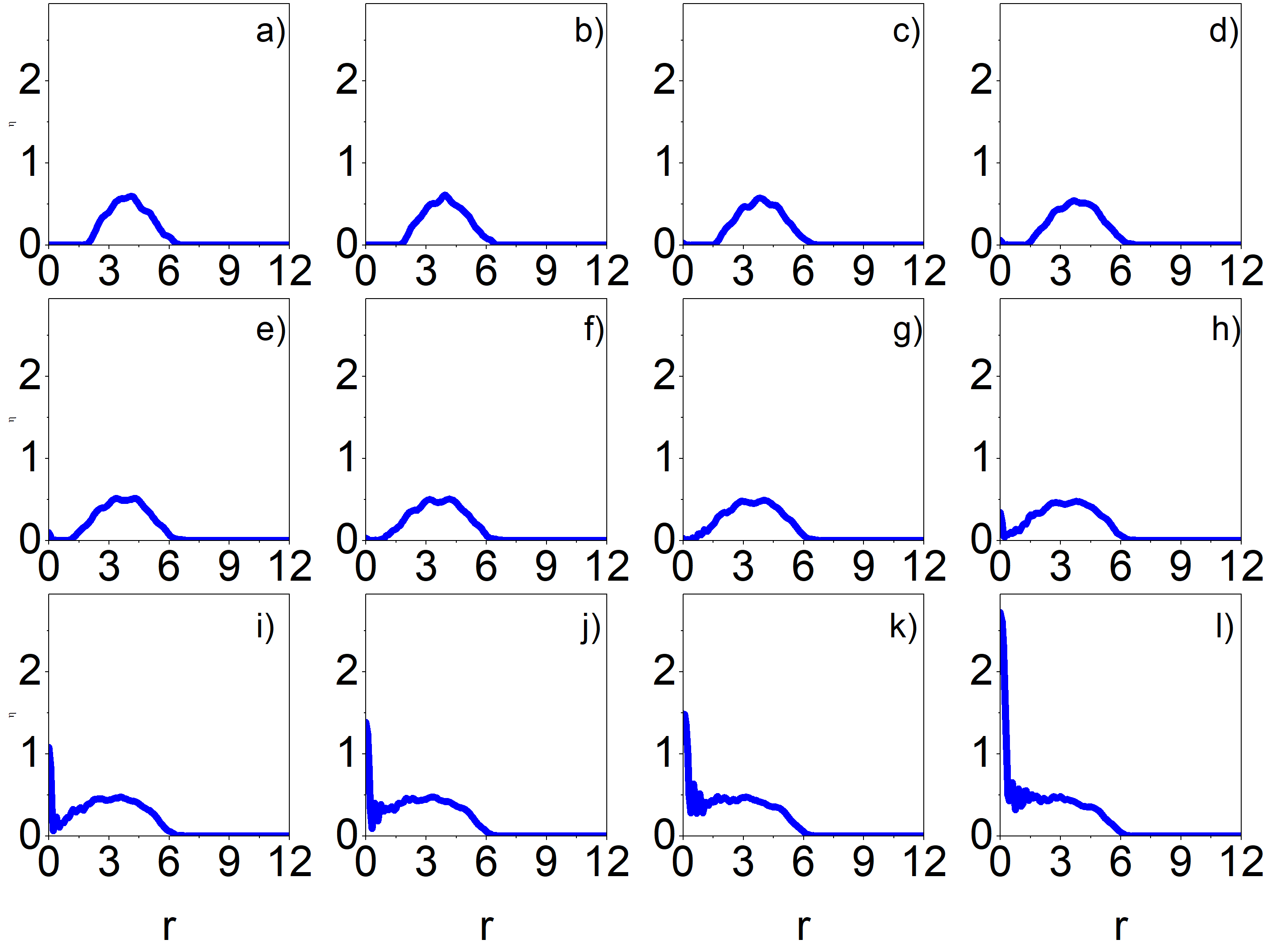}
\caption{Evolution in time of the normalized radial column density profile. A giant vortex with $S_0=25$ is first quasi-adiabatically generated with sequential $S_j=1$ phase imprints. At $t_0=30.3$ the topological quench is activated causing the collapse of the density towards the center. Panel a) is at time $0.1$ after the anti-imprint. The time between the panels is 0.05.}
\label{Fig2}
\end{figure}

To initiate the collapse, after preparing a vortex of total winding \(S_0\), we implement a
topological quench by applying an instantaneous anti-imprint phase mask
\begin{equation}
\Psi(\mathbf{r},t_0^+)=\Psi(\mathbf{r},t_0^-)\exp\!\left[-iS_0\,\theta(x,y)\right],
\label{eq:antiimprint_methods}
\end{equation}
with \(t_0\) the anti-imprint time. This operation quenches the quantized circulation from \(S_0\) to \(0\) in a single step, i.e. it drives the condensate between widely separated topological sectors. 
In hydrodynamic terms, the quench abruptly removes the azimuthal superflow and the associated centrifugal support of the giant vortex. Because the phase is rewritten on a timescale short compared with the hydrodynamic response, the density initially remains annular even though the circulation has been cancelled. The resulting mismatch launches a rapid inward radial flow and a strong ring focusing.

As an example, in Fig.~\ref{Fig2} we show the evolution of the normalized radial density profile $\eta(t)=n(t)/n_0$, with $n_0$ the peak density at $t=0$, for the $S_0=25$ case after the anti-imprint is applied. The ring rapidly collapses toward the trap centre and produces a pronounced central density spike. Because the condensate is suddenly projected into the trivial sector $S=0$, the origin is no longer topologically excluded. The condensate can therefore populate the first regular central mode supported by the trap, which is the ground state of the harmonic oscillator, whose characteristic width is $\ell_0\sim 1$. Driving a large fraction of the density into such a narrow mode generates the observed density spike. In the noninteracting limit, $g=0$, the post-quench focusing transfers the condensate almost entirely into this central mode, while for $a_s=0.034$, the central accumulation is strongly inhibited. This behaviour is consistent with the energetic argument used below to estimate the collapse threshold. We have tested this protocol for different quench times $t_0$, finding the collapse dynamics to be very robust. We have also considered the case in which the harmonic confinement is switched off immediately before the anti-imprint. In that case, the outcome depends on the breathing phase at the quench time $t_0$: if the condensate is in the outward part of the breathing cycle, the subsequent expansion suppresses the collapse, whereas if it is in the inward part, the central focusing remains clearly visible. The breathing phase at the time of the anti-imprint may therefore provide an additional handle for engineering the outcome of the topological quench.

\begin{figure}
\centering
\includegraphics[width=0.45\textwidth]{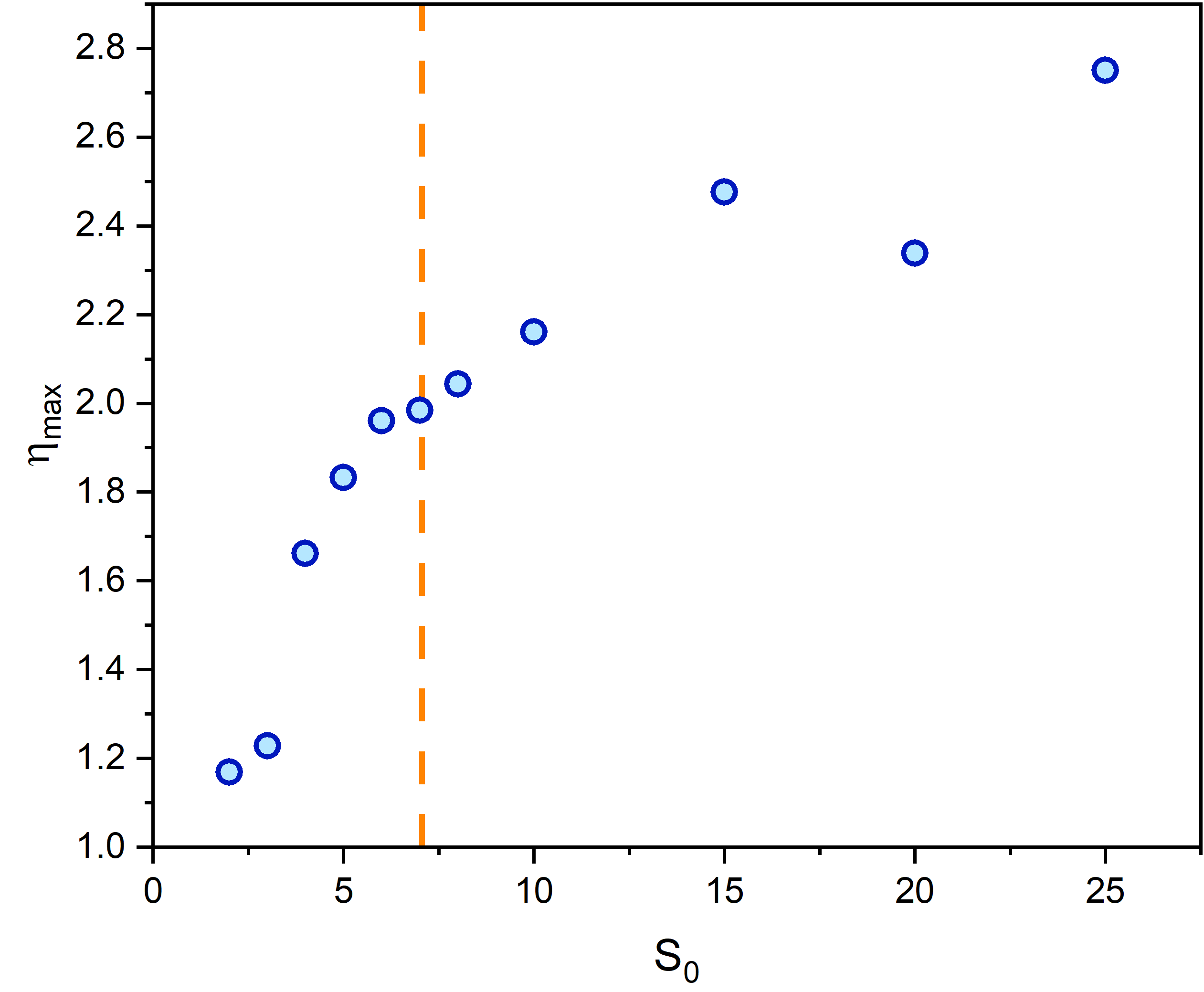}
\caption{Normalized peak density as a function of the initial vortex charge $S_0$. The vertical orange dashed line corresponds to the collapse threshold estimate discussed in the text.}
\label{Fig3}
\end{figure}

\begin{figure}
\centering
\includegraphics[width=0.45\textwidth]{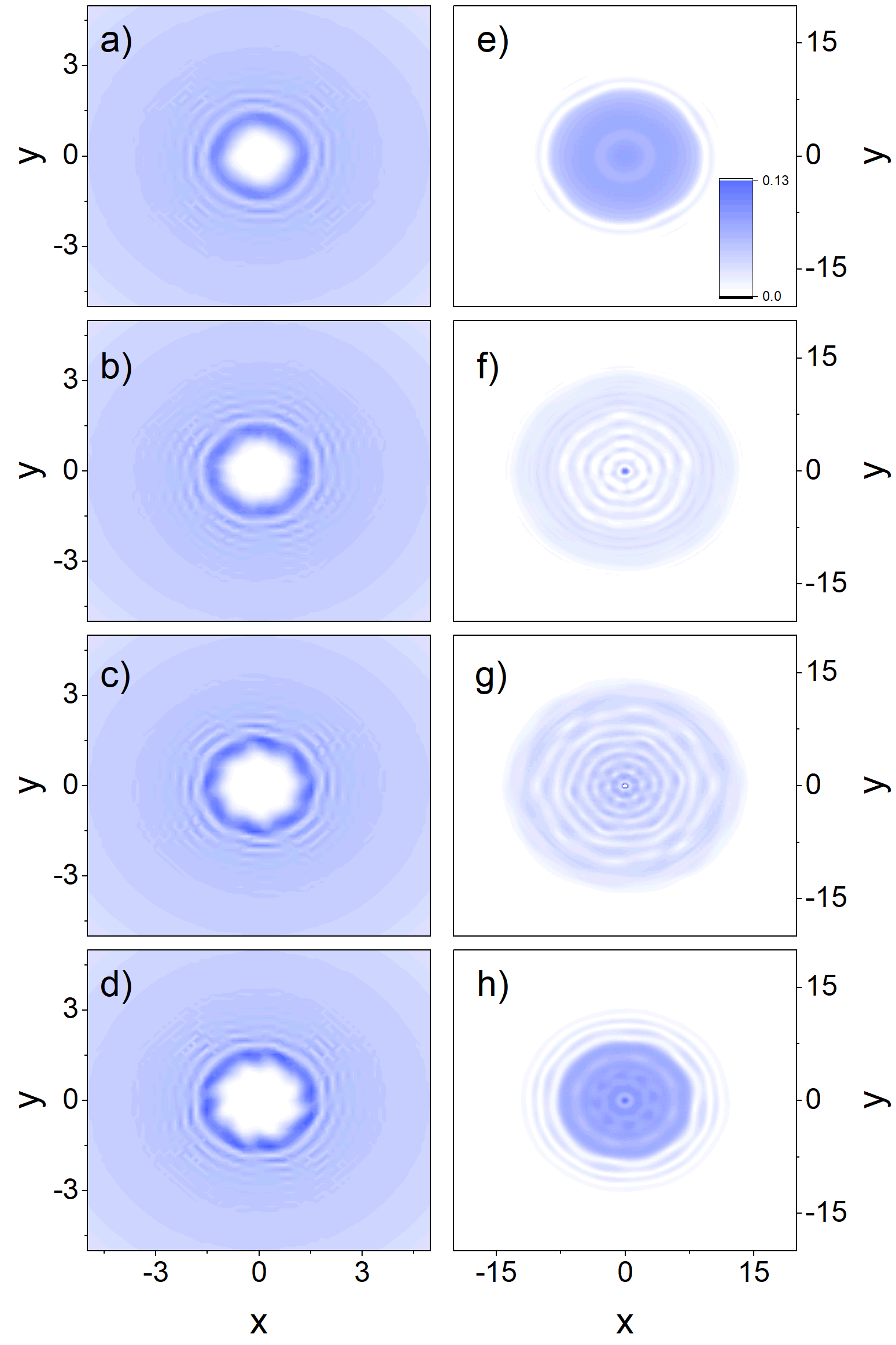}
\caption{\risp{a-d) Column density profiles immediately after the first imprint and e-h) during the collapse after a topological quench starting from $S_0=25$. a) and e) are for a vortex built with an initial imprint of $S_j=4$ and subsequent imprints of $S_j=1$ until $S_0=25$ is reached. b) and f) are for a vortex built with a sequence of $S_j=5, 4,6,3,7$ imprints. c) and g) are for a sequence of $S_j=6,3,4,7,5$ imprints. d) and h) are the same as a) and e) but with an initial imprint of of $S_j=7$. Note the different scales for the left and right columns.}}
\label{Fig4}
\end{figure}

We have applied the topological quench protocol on a range of initial topological charges $S_0$. In Fig. \ref{Fig3} we report the maximum central density (normalized to the central density at $t=0$) reached during the collapse as a function of $S_0$. We observe a step-like change between a small-\(S_0\) regime, where the post-quench increase of the central density is modest, and a large-\(S_0\) regime, where the inward focusing becomes substantial. This can be qualitatively understood with a simple energetic argument, 
that also provides a threshold estimate. By equating the harmonic energy released as the ring of radius \(r_0\) falls to the centre,
\(\varepsilon_{\rm trap}\sim r_0^2/2\),
to the characteristic energy cost of populating a central core of width \(\ell_0\),
\(\varepsilon_{\rm core}(\ell_0)\sim 1/(2\ell_0^2)+\ell_0^2/2\sim 1\).
This gives the condition \(r_0\gtrsim \sqrt{2}\). Using the scaling \(r_0\sim \xi S_0\), we obtain \(S_{0,c}\sim \sqrt{2}/\xi\simeq 7.07\), which is in reasonable agreement with the simulations, as shown in Fig. \ref{Fig3}.

After the first implosion, the condensate enters a post-quench oscillatory regime characterized by repeated refocusing and re-expansion. During this stage, a large fraction of the excess energy released by the topological quench is radiated as circular sound waves, generating a sequence of concentric nonlinear wave fronts. \risp{For the representative case $S_0=25$, the dynamics during the first implosion remains predominantly radial. During the subsequent re-expansion and refocusing, however, the circular fronts can lose cylindrical symmetry and develop pronounced polygonal distortions. The onset of this instability depends sensitively on the imprinting history used to prepare the giant vortex, giving rise to a broad phenomenology with different polygonal structures appearing for different preparation protocols. For example, when the build-up begins with a single large imprint followed by unit steps up to $S_0=25$, the subsequent implosion develops a clear polygonal instability. By contrast, when the vortex is first built through many $S_j=1$ quasi-adiabatic imprints and only the final imprint has large $S_j$, the post-quench dynamics remains essentially axisymmetric. Robust polygonal instabilities are also observed when the entire preparation is performed with large imprinting steps.  As shown in Fig.~4, the origin of this mechanism can indeed be traced to the first imprinting step. During the evolution following the first phase imprint, the vortex core is not perfectly circular, but carries a characteristic polygonal distortion whose symmetry is set by the imprinted charge (for $S_j\geq4$).  This reflects the intrinsic tendency of a multiply quantized vortex to lower its
circulation energy by evolving into separated singly
quantized vortices, whose mutual repulsive interaction favors
a polygonal arrangement. Although this distortion is no longer apparent during the build-up of the larger vortices, as shown in Fig.~1, it remains encoded in the state and is later amplified during the implosive post-quench dynamics. These observations indicate that topological imprints during the build-up stage seed weak azimuthal perturbations, and that the subsequent topological quench acts as a nonlinear amplifier of those hidden angular modes.} In many realizations, the first clearly identifiable pattern is an octagon, which may reflect a residual preference for a fourfold symmetry channel associated with the square numerical grid and the finite simulation box. For longer evolution times, typically after a couple of re-focusing cycles, the polygonal patterns lose coherence and evolve into a disordered state, often accompanied by the emission of Jones--Roberts solitons \cite{meyer} and turbulence. 

In conclusion, we have shown that a repulsive Bose-Einstein condensate can be driven into implosive dynamics by a direct topological quench. By abruptly removing the circulation of a giant vortex, the quench converts stored azimuthal flow into a strong inward focusing, producing a pronounced central density buildup despite the repulsive interactions. The strength of this initial implosion is controlled by the total winding $S_0$, for which we identify a clear threshold for substantial refocusing. Beyond the first implosion, the condensate enters a post-quench oscillatory regime of repeated refocusing and re-expansion, during which circular nonlinear wave fronts are emitted. We find that this later dynamics can undergo a robust breaking of cylindrical symmetry and develop polygonal instabilities. Crucially, this symmetry-breaking channel retains memory of the way the giant vortex was built: the imprinting protocol seeds the azimuthal symmetry content that is later amplified during the post-implosion evolution. Our results introduce topological quenches as a new tool for engineering  implosive dynamics and also their ensuing symmetry-breaking instabilities. More broadly, they establish  quantum fluids as a controllable setting in which selected low-order asymmetries can be programmed, amplified, and studied during implosive evolution.

\paragraph*{Acknowledgements}
We acknowledge fruitful discussions with the members of the Atomic Quantum Systems group at the University of Birmingham, and the use of computing power provided by the Advanced
Research Computing centre at the University of
Birmingham. D.M.G. acknowledges support from the Leverhulme Trust under grant RPG-2024-401. 

\bibliography{main_bibl}

\end{document}